\documentclass[a4paper,UKenglish,cleveref, autoref, thm-restate]{lipics-v2021}

\usepackage{tikz}
\usetikzlibrary{positioning, arrows.meta, shapes.multipart, fit, calc}
\usepackage{listings}
\usepackage{xcolor}
\usepackage{ulem}
\usepackage[utf8]{inputenc}
\usepackage{booktabs}

\lstdefinelanguage{json}{
    basicstyle=\ttfamily\small,
    morestring=[b]",
    stringstyle=\color{blue},
    morecomment=[l]{(*},
    commentstyle=\color{gray},
    morecomment=[l]{*)},
}

\title{LogicProof: An Interactive Web-Based Educational Theorem Prover for Natural Deduction and Sequent Calculus across Classical and Constructive Logics} 

\titlerunning{LogicProof: An Interactive Web-Based Educational Theorem Prover ...} 

\author{Ján Perháč\footnote{Corresponding author}}{Department of Computers and Informatics,
Faculty of Electrical Engineering and Informatics, 
Technical University of Košice,
Letná 9, 042 00 Košice,
Slovak Republic \and \url{https://kpi.fei.tuke.sk/en/person/jan-perhac}}{jan.perhac@tuke.sk}{https://orcid.org/0000-0001-6347-2409}{}

\author{Vasyl Khashcha}{Department of Computers and Informatics,
Faculty of Electrical Engineering and Informatics, 
Technical University of Košice,
Letná 9, 042 00 Košice,
Slovak Republic \and \url{https://www.linkedin.com/in/vasylkhashcha?utm_source=share_via&utm_content=profile&utm_medium=member_ios}}{vasyl.khashcha@student.tuke.sk}{https://orcid.org/0009-0001-2400-671X}{}

\author{Samuel Novotný}{Department of Computers and Informatics,
Faculty of Electrical Engineering and Informatics, 
Technical University of Košice,
Letná 9, 042 00 Košice,
Slovak Republic \and \url{https://kpi.fei.tuke.sk/en/user/11125}}{samuel.novotny@tuke.sk}{https://orcid.org/0009-0005-1675-2965}{}

\authorrunning{J. Open Access and J.\,R. Public} 

\Copyright{Jane Open Access and Joan R. Public} 



\category{} 

\relatedversion{} 

\funding{This research was supported by the Cultural and Educational Grant Agency (KEGA) under project No. 052TUKE-4/2025, "Modern Approaches in Educating IT Professionals in the Field of Type Theory".}

\nolinenumbers 

\EventEditors{John Q. Open and Joan R. Access}
\EventNoEds{2}
\EventLongTitle{Tools for Educational Activities in Logic (TEAL 2026)}
\EventShortTitle{TEAL 2026}
\EventAcronym{TEAL}
\EventYear{2026}
\EventDate{July 25, 2026}
\EventLocation{Lisbon, Portugal}
\EventLogo{}
\SeriesVolume{42}
\ArticleNo{23}

\begin{document}

\maketitle

\begin{abstract}
We present \textit{LogicProof}, an interactive web-based theorem prover designed for educational use. The system supports natural deduction and sequent calculus for propositional and first-order logic in both classical and constructive variants. It emphasizes a modern user experience through real-time feedback and interactive visualization of proof trees. We evaluated \textit{LogicProof} in a study involving 35 students. The results suggest that the system improves understanding of formal proof construction and supports student engagement with logical concepts. Compared to traditional pen-and-paper approaches, students reported faster iteration, easier error correction, and greater confidence in the correctness of their solutions. These findings indicate that \textit{LogicProof} can serve as a practical supplement to existing teaching methods, particularly in supporting the learning process and reducing barriers associated with formal proof construction.
\end{abstract}

\section{Introduction}
\label{sec:intro}

The traditional teaching approach based on manual problem-solving tasks on paper has notable limitations, most importantly the absence of immediate feedback. When students construct proofs on paper, they often recognize logical errors only after evaluation, which slows down the learning process. 

Existing professional software solutions address this issue only partially \textit{Full-scale Interactive Theorem Provers (ITPs)}, such as Rocq (\textit{Coq}) or Lean are highly expressive and powerful but come with a steep learning curve that can overwhelm beginners. 

On the other hand, many existing educational tools, such as \textit{Natural Deduction Proof Generator \& Checker} \cite{ndprover} and \textit{Natural Deduction Tool} \cite{natural_deduction_tool}, often behave like a simple "\textit{calculator}": they produce a proof as output from a given input formula without explaining the intermediate steps.

Other tools are often built on top of full-scale proof assistants, such as \textit{Rocq Game} \cite{rocq-game}, \textit{ProofBuddy}, or \textit{Lean Natural Number Game} \cite{lean-game-server}. They introduce more a user friendly environment, or gamified approaches. Nevertheless, these solutions still inherit many of the underlying complexities of the original systems.

Another category of educational tools, such as \textit{OnlineProver} \cite{perhac2024onlineprover} and \textit{Proof-tree-builder} \cite{Korkut2023}, attempts to simulate pen-and-paper proof construction; however, their input mechanisms and overall design are often not user-friendly. 

Across all categories, common limitations include limited user interfaces, unintuitive or complicated input syntax, and support only a specific proof formalism.

Our tool presented in this paper, the \textit{LogicProof} (\textit{avaliable online at} \cite{logicProof-app}) fits into the last category. The system provides an interactive environment for proof construction using natural deduction and sequent calculus. It employs a Gentzen-style and Fitch-style representation. The application supports propositional logic, first-order logic (FOL), extensions such as Robinson arithmetic and linear order theory \cite{tarski1953, logika}.

The core philosophy of \textit{LogicProof} is to guide students through the proof construction process. The tool validates the application of inference rules step by step in real time and visualizes proofs as interactive proof trees, allowing students to focus on the logical structure rather than syntactic details.

To ensure the effectiveness of the tool as an educational aid, its development was guided by several key objectives:
\begin{itemize}
\item \textit{Implementation of a logical core:} We developed a high-performance parser for first-order logic formulas and implemented algorithms for verifying deduction rules. A key technical challenge was the correct implementation of quantifier rules, including substitution constraints and variable scope management.
\item \textit{Support for formal theories:} The system was extended to support axiomatic theories \cite{tarski1953, logika}, allowing users to incorporate axioms alongside assumptions.
\item \textit{Interactive User Interface:} We designed a graphical user interface (GUI) that represents proofs as interactive trees, helping students internalize the hierarchical structure of formal reasoning.
\item \textit{Accessibility and Deployment:} The tool is implemented as a platform-independent single-page application (SPA), ensuring easy integration into curricula and immediate accessibility without requiring installation.
\item \textit{Modern input method:} We implemented a method to input logical formul\ae~ with the same syntax as used on a blackboard. 
\end{itemize}

This paper is organized as follows. Section \ref{sec:tool} provides a technical overview of the architecture and implementation of the system, 
and its functionalities. Section \ref{sec:evaluation} presents the results of a user study involving 35 university students, which evaluated both pedagogical effectiveness and the usability. Section \ref{sec:related} reviews existing logic pedagogical tools and places \textit{LogicProof} within the state of the art. Finally, section \ref{sec:conclusion} summarizes the contributions and outlines directions for future work.

\section{The Tool}
\label{sec:tool}
In this section, we analyze the system requirements, architectural decisions, the implementation of the logical core and functionalities of the tool. Based on these requirements, we identified several key objectives.

The system's functionality is centered on four primary pillars:
\begin{itemize}
    \item \textbf{Syntactic Analysis:} Support for FOL syntax, including n-ary predicates, nested terms, and quantifiers with proper scope management.
    \item \textbf{Proof Validation:} Real-time validation of the derivation rules application.
    \item \textbf{Theory Integration:} Capability to incorporate  axioms, specifically providing a complete environment for first order theories.
    \item \textbf{Proof Visualization:} An interactive tree-based representation that allows users to create and manipulate formal proofs.
\end{itemize}

Non-functional requirements focused on \textbf{accessibility}, ensuring the tool is a zero-install web application; \textbf{usability}, minimizing the cognitive load during complex proof steps; and \textbf{extensibility}, utilizing a modular architecture to support future additions of various logical systems.

\subsection{Technology Stack}

An important design decision was to not use heavy frontend frameworks (such as React or Angular) in favor of a \textbf{"Vanilla JS" stack}. This approach was motivated by the specific performance requirements of rendering large, recursive proof trees. By directly manipulating the Document Object Model (DOM) and avoiding the overhead of a Virtual DOM diffing engine, we achieved significantly faster re-rendering times during complex tree transformations \cite{zakas2010, strazzullo2019}.

\begin{itemize}
\item \textbf{HTML5 (HyperText Markup Language):}
Used to define the semantic structure of the application. Modern HTML5 elements ensure accessibility and proper document hierarchy. In our application, HTML primarily serves as a container for dynamically generated SVG elements representing proof trees, as well as for editor controls.

\item \textbf{CSS3 (Cascading Style Sheets):}  
Responsible for visual presentation and layout. We utilize modern techniques such as \textit{CSS Grid} and \textit{Flexbox} to achieve a responsive design, eliminating the need for external CSS frameworks (e.g., Bootstrap). The modularity of styles enables easy customization of visual themes (e.g., dark mode) without modifying application logic.

\item \textbf{JavaScript (ES6+):}  
Serves as the primary programming language for application logic, using modern ECMAScript features (modules, classes, arrow functions).

\item \textbf{D3.js:}  
A data visualization library used for generating syntactic trees of logical formulas. D3.js provides efficient algorithms for hierarchical layout computation and supports smooth transformations (zooming and panning) within SVG containers.

\item \textbf{Monaco Editor:}  
Integrated as the input interface for logical formulas (the core editor of VS Code). It provides a professional editing environment with syntax highlighting, intelligent code completion, and real-time error visualization, significantly improving the user experience (UX).

\item \textbf{Driver.js:}  
A lightweight library used to implement an interactive guided tour of the application. It facilitates intuitive user tutorial by visually highlighting and explaining key interface components during the initial use.

\item \textbf{Webpack and Babel:}  
Although the application is based on “vanilla” JavaScript, modern tooling is used for module management and optimization. \textit{Webpack} handles bundling and dependency management, while \textit{Babel} ensures backward compatibility through transpilation for older browsers.

\item \textbf{ANTLR4 (TypeScript target):}  
A tool for generating lexers and parsers, enabling us to formally define the grammar of logical expressions and automatically generate processing code. While the rest of the application is written in JavaScript, the parser leverages generated typed interfaces to ensure correctness and robustness of input processing.

\end{itemize}

\subsection{Client-Side Architecture and Build Process}
The application is designed following the Single Page Application paradigm. All application logic, including formula parsing and proof tree validation, is executed directly in the user's browser (client-side). This approach eliminates the need for continuous server communication at each proof step, ensuring low latency and immediate responsiveness of the user interface.

Although the application is written in standard \textbf{JavaScript (ES6+)}, we employ \textbf{Webpack} for code organization and preparation for production deployment. The project is not implemented as a monolithic script but is instead logically divided into multiple independent modules (e.g., Logic Core, UI Handlers, ANTLR definitions).

Within our architecture, Webpack fulfills several key roles:
\begin{itemize}
\item \textbf{Module Bundling:} It combines application logic and automatically generated ANTLR files (e.g., GrammarLexer.js, GrammarParser.js, GrammarListener.js) into a single optimized bundle (bundle.js), reducing overhead associated with loading multiple small files over the network.
\item \textbf{Dependency Resolution:} It ensures the correct loading order of modules. This is particularly critical for inheritance relationships, where our class \texttt{MyGrammarListener} extends the generated \texttt{GrammarListener}. Webpack guarantees that the parent class is available before the child is initialized.

\item \textbf{Code optimization:} For production builds, Webpack removes whitespace and comments and shortens internal variable names, reducing file size and improving load times.
\end{itemize}
The result of this process is a static artifact that can be deployed on any web server without requiring backend-specific configuration.

\subsection{The User Interface}
\begin{figure}[ht]
    \centering
    \includegraphics[width=0.99\linewidth]{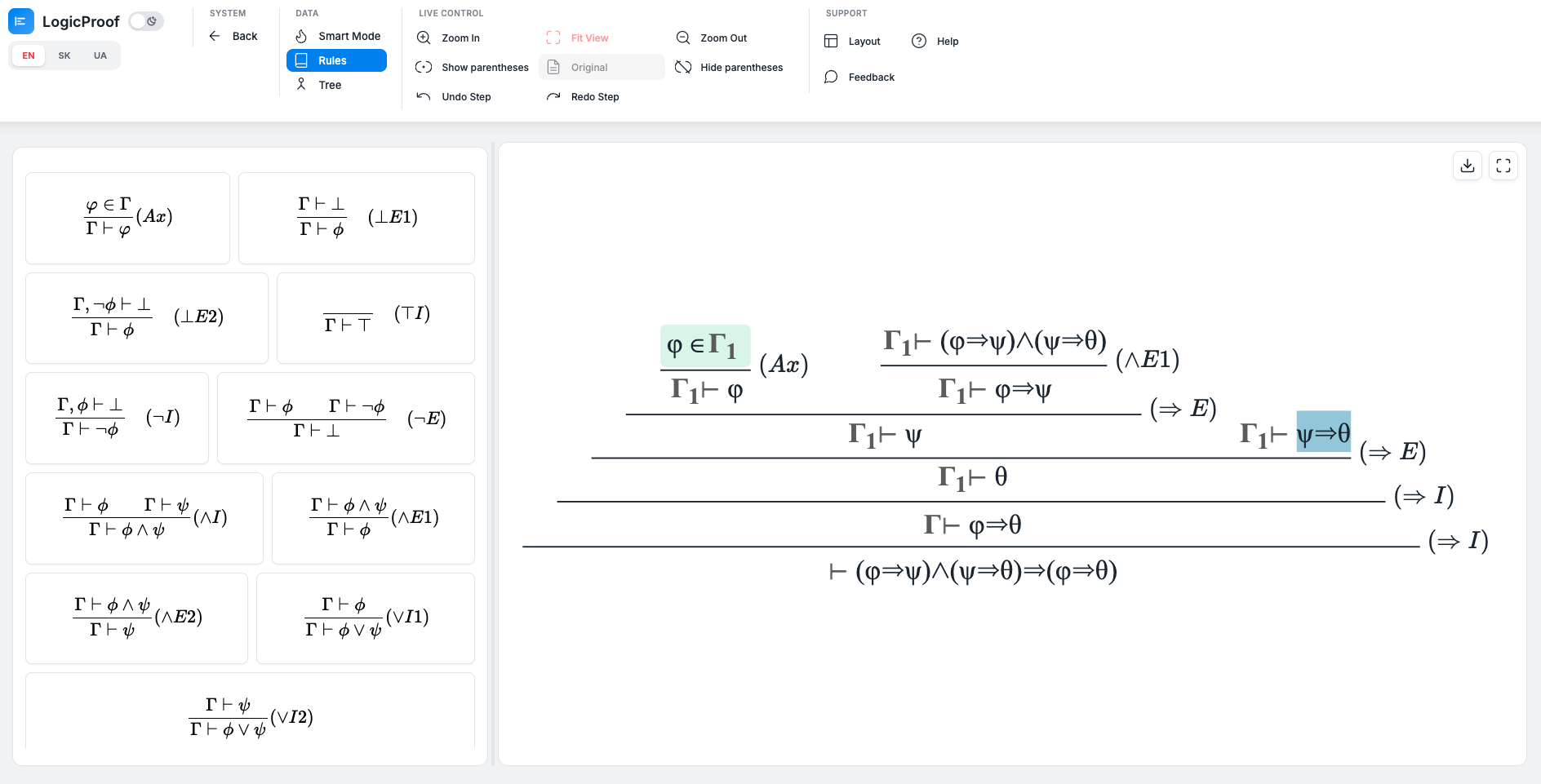}
    \caption{Main window of the tool.}
    \label{fig:mainwindow}
\end{figure}

The UI is designed to simulate a professional Integrated Development Environment (IDE) while remaining accessible to novices. The main window with proof is depicted in Figure \ref{fig:mainwindow}.

It is divided into three primary sections, all of which are fully resizable and allow flexible layout customization (e.g., by rearranging the position of each section).

\begin{itemize}
    \item \textbf{Control Panel:} This section allows the user to configure the application and manipulate the proof process. It provides the following functionalities:
    \begin{itemize}
        \item Switching between color themes (light and dark mode).
        \item Changing the application language. The tool currently supports three languages: English, Slovak, and Ukrainian.
        \item A \textit{Back} button that enables navigation within the application, such as returning to the \textit{welcome screen}, clearing the current proof to start a new one, or correcting the input formula.
        \item Switching between \textit{Smart Mode} and the standard mode for rule suggestion, as well as displaying the abstract syntax tree of the currently selected node in the proof.
        \item Tools for proof manipulation, including zooming in and out, stepping backward and forward, and toggling the visibility of parentheses.
        \item Access to help resources, including documentation, layout configuration options, and the ability to submit anonymous feedback.
    \end{itemize}

    \item \textbf{Rule List:} This section displays the available deduction rules. Depending on the selected mode, it either shows all rules or only those applicable to the current state of the proof.

    \item \textbf{Proof Window:} This section visualizes the current state of the proof. The user interacts with the proof by selecting a node and then choosing an appropriate rule from the rule list. Additionally, the user can explore the contents of the \textit{context $\Gamma$} by clicking on it.
\end{itemize}

\begin{figure}[ht]
    \centering
    \includegraphics[width=0.8\linewidth]{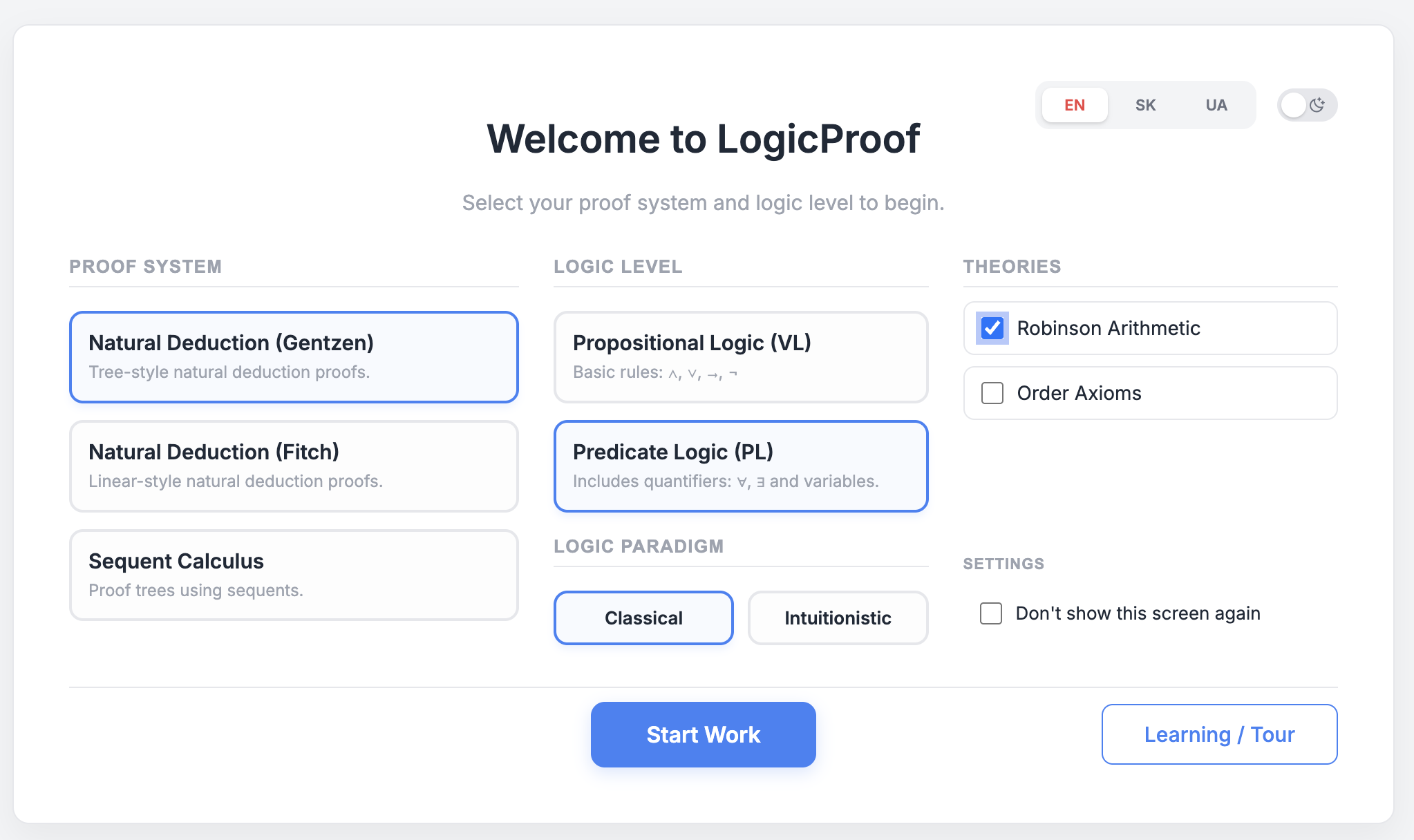}
    \caption{Welcome screen of the tool.}
    \label{fig:welcome}
\end{figure}

The welcome screen, depicted in Figure~\ref{fig:welcome}, provides options for the initial configuration of the application. 

\noindent
The user can switch the language, toggle between light and dark themes, and select the logic, its paradigm, and the proof system. Additionally, the user may disable the welcome screen for future visits. An important feature is the \textit{Learning/Tour} mode, which provides an interactive tutorial guiding the user through the application.

\begin{figure}[ht]
    \centering
    \includegraphics[width=0.7\linewidth]{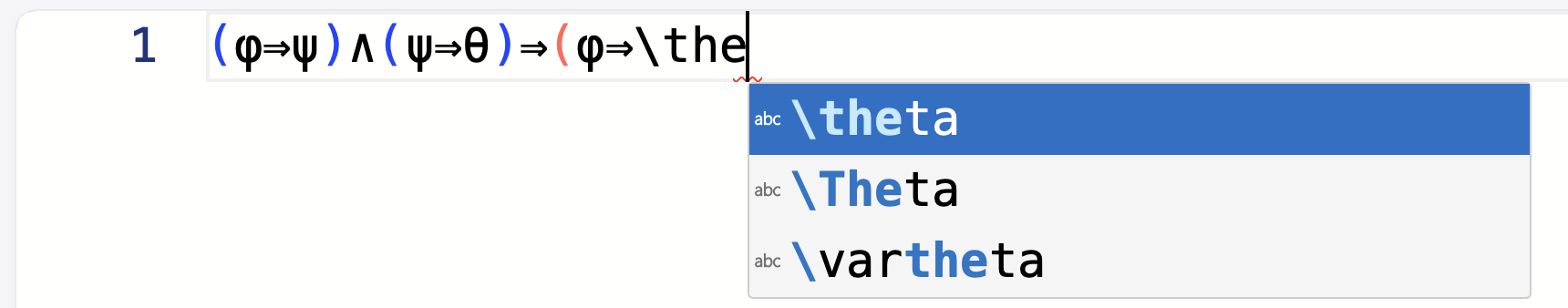}
    \caption{Input of a formula.}
    \label{fig:editor_input}
\end{figure}

We integrated the \textbf{Monaco Editor} to provide a high quality editing experience. We implemented a custom \textbf{Monarch Tokenizer} that provides real-time syntax highlighting for logical symbols. An example is depicted in fugure \ref{fig:editor_input} 

\begin{figure}[ht]
    \centering
    \includegraphics[width=0.8\linewidth]{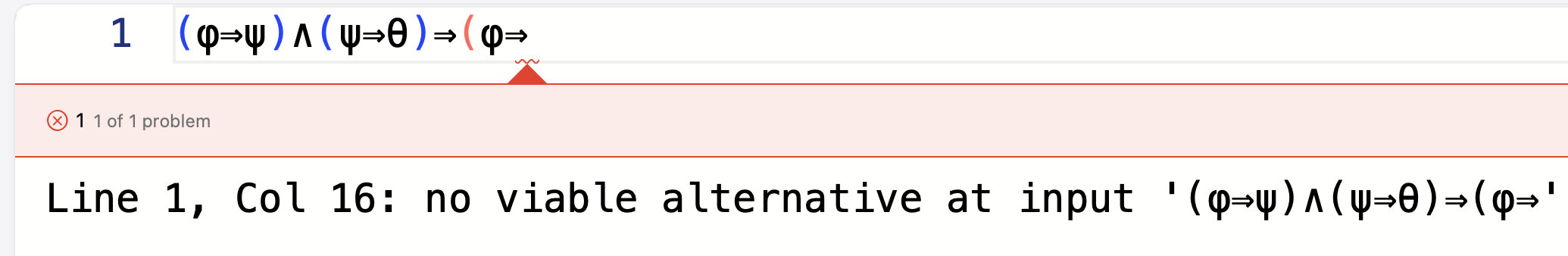}
    \caption{An syntactic error.}
    \label{fig:editor_error}
\end{figure}

Furthermore, the editor is linked directly to the ANTLR4 parser, allowing it to provide "as-you-type" feedback. Syntactic errors are underlined, and hover-tooltips provide explanations of the required syntax (Figure \ref{fig:editor_error}).

For parsing the input we have used the \textbf{ANTLR} parser generator. 
We defined a our own formal grammar for FOL and Robinson Arithmetic \cite{aho2007compilers, huth2004logic}. A fragment of the grammar is depicted in Figure \ref{lst:logic_precedence_en}. The whole grammar is published in the GitHub repository \cite{logicProof-repo}.


\begin{figure}[h]
\begin{lstlisting}[frame=single,numbers=left,basicstyle=\footnotesize\ttfamily]
// Recursive descent for operator precedence
implication: disjunction (IMPL implication)?;
disjunction: conjunction (DIS conjunction)*;
conjunction: negation (CON negation)*;
negation: NEG+ quantified | quantified;
\end{lstlisting}
\caption{ANTLR4 definition for logical operator precedence.}
\label{lst:logic_precedence_en}
\end{figure}

Proof validation is performed through a three-layer pipeline:
\begin{enumerate}
    \item \textbf{Syntactic Layer:} Ensures the formula is well-formed according to the grammar.
    \item \textbf{Semantic Layer (Rule Handlers):} Each inference rule is defined as a declarative handler. These handlers verify if the rule is applicable to the current AST structure (e.g., checking for a conjunction before applying $\land E$).
        \item \textbf{Contextual Layer (Scope \& Unification):} The system verifies "Eigenvariable" conditions for quantifier rules ($\forall I, \exists E$) \cite{vanDalen2013} by recursively checking for free variables in open hypotheses.
\end{enumerate}

\section{Evaluation}
\label{sec:evaluation}

To assess the pedagogical effectiveness and usability of \textit{LogicProof}, we conducted an experimental evaluation. The study focused on the system's intuitiveness, ease of use, and overall user experience (UX), with a particular emphasis on comparing the process of constructing formal proofs on paper versus using the tool.

\subsection{Methodology and Participants}

The tool was developed during the academic year 2025/2026. In the spring semester, we introduced it to students enrolled in the Master's-level course \textit{Logic for Informaticians} at the Technical University of Košice. The study involved \textbf{35 participants}, all of whom were students in this course. This ensured that participants had sufficient background knowledge to meaningfully evaluate the system.

The evaluation questionnaire consisted of three main parts:
\begin{enumerate}
\item \textbf{Practical User Scenarios:} Task-based evaluations designed to assess the usability and correctness of core features under realistic conditions.
\item \textbf{System Usability Scale (SUS):} A standardized usability measure consisting of 10 items rated on a 5-point Likert scale \cite{sauro2011sus}.

\item \textbf{Qualitative Feedback:} Open-ended questions comparing \textit{LogicProof} with the traditional pen-and-paper approach, as well as collecting feedback on the tool itself, such as identified issues and suggestions for improvement.

\end{enumerate}

The first question collected information about the time spent using the tool. Approximately 45.7\% of participants reported spending between 15 and 30 minutes using the application, while 28.6\% spent 30 to 60 minutes. A smaller group (8.5\%) used the tool for more than one hour, while 17.1\% reported spending less than 15 minutes. This suggests that most participants interacted with the system long enough to form informed opinions.

\subsection{Quantitative Results: SUS Analysis}

The SUS score \cite{sauro2011sus} achieved by \textit{LogicProof} was \textbf{79.5 out of 100}. According to standard interpretations, a score of 68 represents the global average; therefore, a score of 79.5 places the system in the \textbf{"Good" to "Excellent"} category (Grade A-). This indicates that the system is highly usable for its target audience and does not suffer from major usability issues. After applying a 10\% trimmed mean, the score increased to \textbf{81.12}, corresponding to a \textbf{Grade A}.

Analysis of individual responses in the SUS questionnaire further revealed several key strengths:
\begin{itemize}
\item \textbf{High Functional Integration:} Most respondents agreed that the system’s features are well integrated and work together consistently.
\item \textbf{Autonomy and Confidence:} Users felt confident using the application and indicated that they would not require additional technical support.
\item \textbf{High Acceptance Rate:} Approximately 83\% of participants rated the tool’s usability as above average.
\end{itemize}

\subsection{Scenario-Based Performance}

Two practical scenarios were designed to evaluate the core components of the application:
\begin{itemize}
\item \textbf{Scenario 1 (Basic Interaction):} Enter a formula and construct a proof. \ \textbf{Success rate: 91\% (32/35)}. The main issue identified was initial confusion regarding the syntax of logical operators.
\item \textbf{Scenario 2 (Complex Rules and Branching):} Apply rules that lead to proof branching. \ \textbf{Success rate: 83\% (29/35)}. The primary issue was the occasional overlap of modal windows with the proof tree, which obscured the context.
\end{itemize}

\subsection{Pen and Paper Proofs vs. LogicProof}

The next part of the questionnaire consisted of the following questions:
\begin{itemize}
\item What do you consider to be the advantages of constructing proofs on paper compared to using \textit{LogicProof}?
\item What do you consider to be the advantages of constructing proofs using \textit{LogicProof} compared to the pen-and-paper approach?
\end{itemize}

The responses suggest that students perceive both approaches as having distinct benefits, often depending on personal preference and experience.

Approximately 40–45\% of students reported no clear advantage of solving tasks on paper compared to the application. Among the remaining responses, around 30–35\% highlighted better memorization and deeper cognitive engagement when writing by hand. About 20–25\% emphasized greater flexibility and control, including freedom in notation and the ability to structure proofs without strict syntactic constraints. A smaller group (10–15\%) mentioned having a better overview of proofs on paper. However, several students also noted that this approach can be slower and less efficient.

In contrast, \textit{LogicProof} was perceived as more efficient and supportive. Around 50–60\% of students emphasized speed and ease of use, particularly the ability to correct mistakes and iterate quickly. Approximately 40–50\% highlighted automatic validation and correctness checking as a major benefit. Features such as undo, history, and smart hints were mentioned by 30–40\% of respondents as helpful support mechanisms. Additionally, around 25–35\% appreciated the clear and structured visualization of proofs.

\subsection{User Feedback Analysis}

The final part of the questionnaire focused on collecting feedback about the tool. We have asked the following questions:
\begin{itemize}
    \item Please describe a specific issue you encountered while using the application that caused you the most difficulty. How did you resolve it?
    \item Do you have any suggestions that would improve the quality of this tool for you?
    \item Do you have any additional comments, observations, or did you encounter any bugs?
\end{itemize}
The responses were categorized into three main areas: encountered issues, problem-solving strategies, and suggestions for improvement.

\subsubsection{Encountered Issues}
Approximately \textbf{30\%} of respondents reported \textbf{no significant problems} while using the application, indicating generally good baseline usability. However, several recurring issues were identified:

\begin{itemize}
    \item \textbf{User Interface and Navigation (25\%):} Users reported difficulties with unintuitive UI behavior, such as unclear next steps after rule selection, confusion during initial interaction, and problems locating features (e.g., \LaTeX{} export).
    \item \textbf{Understanding the Proof Process (20\%):} A significant portion of users struggled with understanding how to begin a proof or correctly apply deduction rules, particularly in natural deduction.
    \item \textbf{Symbol Input and Notation (15\%):} Users found it difficult to input logical symbols (e.g., $\lor$, $\land$) and were unfamiliar with the notation.
    \item \textbf{Tree Visualization and Layout (5\%):} Some users experienced issues with large proof trees not fitting on the screen or being obscured by modal dialogs.
    \item \textbf{Bugs and Technical Issues (5\%):} Reported issues included inconsistent behavior of the “step back” function, occasional freezing in sequent calculus, and visual ambiguity between symbols such as $\phi$ and $\varphi$.
\end{itemize}

\subsubsection{Problem-Solving Strategies}
The most common approaches to resolving issues were:

\begin{itemize}
    \item \textbf{Trial and Error (50\%):} The majority of users relied on experimentation to understand the system.
    \item \textbf{Peer Assistance (15\%):} Some users consulted classmates or relied on prior demonstrations.
    \item \textbf{Interface Exploration (15\%):} Users explored UI elements (buttons, menus) to discover functionality.
    \item \textbf{Reset/Reload (10\%):} In cases of errors or dead ends, users refreshed the application.
    \item \textbf{No Action Needed (10\%):} Users who encountered no issues did not require any specific strategy.
\end{itemize}

\subsubsection{Suggestions for Improvement}
User suggestions highlight several areas for enhancement:

\begin{itemize}
    \item \textbf{Syntax Input (25\%):} Better explanation how to input special symbols or adding shortcuts for logical operators.
    \item \textbf{Better Onboarding (20\%):} Inclusion of a tutorial or interactive guide for first-time users.
    \item \textbf{UI/UX Enhancements (20\%):} Improving clarity of controls, rule selection, and feature visibility (e.g., \LaTeX{} export).
    \item \textbf{Editing and Undo Features (15\%):} More robust undo/reset functionality and finer control over proof tree editing.
    \item \textbf{Performance and Visualization (10\%):} Better handling of large proof trees and improved navigation (e.g., zooming, panning).
    \item \textbf{Advanced Features (10\%):} Requests included searchable rule selection and optional automated solving hints.
\end{itemize}

To address these issues, we enhanced the in-tool documentation, introduced an interactive tutorial that systematically guides users through the main functionalities, and resolved the reported issues related to navigation within the proof environment. While some users suggested automated proof generation, such functionality is beyond the scope of this work, as the primary objective is to support learning through interactive, user-driven proof construction. An auto-focus feature was introduced to automatically select the next open premise after applying a rule, improving efficiency.

\subsubsection{Additional Comments and Bugs}
Most users (\textbf{60\%}) reported no additional issues, confirming overall system stability. Reported concerns include:
\begin{itemize}
    \item Minor UI inconsistencies (e.g., language mismatches, unclear labels such as “Smart Mode”).
    \item Loss of progress upon page refresh without warning.
    \item Occasional rendering or navigation glitches.
    \item Lack of clarity in certain features (e.g., tree visualization or input format requirements).
\end{itemize}

In response to the identified issues, we have addressed the reported UI inconsistencies by improving label clarity and ensuring consistency across supported languages. To minimize the loss of progress upon page refresh, we introduced a navigation mechanism that allows users to seamlessly move within the application and choose whether to retain the current formula or start with a new one.

\section{Related Work}
\label{sec:related} 

The development of educational software for formal logic is an very active area of research and has a rich history, evolving from early terminal-based systems to modern interactive web applications. This section reviews the existing landscape, categorizing tools by their pedagogical focus, proof formalisms, and technical capabilities.

State-of-the-art proof assistants such as \textit{Rocq (formerly Coq)} \cite{coq, bertot2013}, \textit{Lean}, and \textit{Isabelle/HOL} are primarily used to prove complex mathematical theorems and verify the correctness of critical software. Their interaction model---tactic-based scripting---is often described in our analysis as ``\textit{too complex for beginners}.'' These tools require students to learn a specialized programming language before they can prove even simple logical tautologies. While some educational tools built on top of them, such as \textit{Rocq Game} \cite{lean-game-server}, or \cite{karsten2025proofbuddy}, attempt to minimize this issue, they remain fundamentally tied to the underlying complex type theory, which may not align with the goals of introductory logic courses.

Tools such as \textit{Natural deduction prover} \cite{ndprover, broda2006friendly}  are dedicated to the visualization of \textit{Fitch-style} proofs with nested box notation. This style is highly intuitive for propositional logic but becomes increasingly difficult to manage when dealing with first-order logic (FOL) quantifiers and deeply nested assumptions. Tree-based systems, such as \textit{panda} \cite{gasquetEtal2011panda} and \textit{OnlineProver} \cite{perhac2024onlineprover}, represent proofs using the Gentzen style format \cite{gentzen1969}. This visualization is elegant and clearly illustrates the hierarchical structure of dependencies. However, many existing tree-based tools are limited in their interactivity and rely on unintuitive input methods, often using syntax that differs from what is taught in class. These issues can confuse new learners, who are immediately exposed to unfamiliar notation in a language they have only just begun to learn. To address this, \textit{LogicProof} employs modern technologies that enable input and manipulation of formul\ae~using the same notation as on a blackboard.

Despite its important role in proof theory and its utility in computer science, \textit{Sequent Calculus} is rarely supported in lightweight educational tools. The tool \textit{Proof Tree Builder} \cite{Korkut2023} does support sequent calculus; however, it suffers from a complicated input method and a confusing user interface. Most existing provers focus exclusively on natural deduction.
Our tool, \textit{LogicProof}, addresses this limitation by leveraging a modern input editor combined with a simple and clear user interface, enabling students to explore the deeper symmetries of logic that are less apparent in natural deduction.

Another limitation identified in existing educational software is the lack of support for \textit{Constructive Logics}. Most introductory tools, such as \cite{teachinglogic-dn} or \cite{Korkut2023}, are hardcoded for classical logic, assuming the Law of Excluded Middle and Double Negation Elimination by default. To address this issue, our tool allows users to switch between classical and constructive modes. This provides a valuable educational opportunity to observe which formul\ae~are not provable in a constructive setting.
 
The \textit{LogicLab} \cite{sieg2007apros} is a standalone desktop application where students can construct proofs in natural deduction using the Fitch-style notation. Interesting feature of the application is the \textit{Proof Tutor} \cite{perkins2007strategic}, which provides hints and feedback to guide students through the proof process. However, it is not web-based and requires installation, which can be a barrier for some users. 

Different type of tools focus on tutoring features, such as providing hints, feedback, and step-by-step guidance to help students learn and understand logical concepts. The expert system presented in \cite{miwa2014intelligent} is an example of a tool that provides variable levels of instructional support for natural deduction. Similarly, the \textit{LOGAX} tool provides a feature for generating hints and feedback in a Hilbert-style axiomatic proof tutor \cite{lodder2021generation}. These tools aim to enhance the learning experience by offering personalized guidance, but they often lack the flexibility and interactivity of more modern web-based applications. While \textit{LogicProof} does not currently include a dedicated tutoring engine, integrating adaptive tutoring and richer feedback is a promising direction for future work.

Our analysis identified three main areas in which current tools fall short:
\begin{enumerate}
\item \textbf{Syntactic Rigor vs. User Experience:} Many tools either enforce rigid, difficult-to-use syntax or provide unintuitive input methods. There is a lack of tools that incorporate IDE-like features (e.g., real-time syntax highlighting).
\item \textbf{Integration of Multiple Proof Systems:} Many tools are limited to a single deduction calculus or proof style (e.g., Gentzen or Fitch).
\item \textbf{Academic Portability:} Most tools do not allow students to easily export their work into professional formats such as \LaTeX{} for inclusion in assignments or research papers.
\end{enumerate}

To address these issues, we have integrated a modern input editor into our tool, supporting the entry and manipulation of special symbols. We have implemented two proof calculi---natural deduction and sequent calculus---with the ability to switch between classical and constructive logic, along with support for exporting proofs as \LaTeX{} code.

\section{Conclusion}
\label{sec:conclusion}

In this paper, we introduced \textbf{LogicProof}, an interactive web-based tool designed to support the teaching of formal logic in higher education. The goal was to create an environment that balances formal rigor with usability. By combining a lightweight architecture with advanced editing features and a solid logical foundation, the tool makes working with formal proofs more accessible to students. 

The implementation covers key aspects of First-Order Logic, such as quantifiers, substitution, and selected formal theories, including Robinson Arithmetic. It supports both Natural Deduction and Sequent Calculus, in classical as well as constructive settings. This combination is not commonly available in existing lightweight educational tools. The use of a Vanilla JavaScript stack also helped ensure a responsive and smooth user experience, which is important when working with larger proof structures.

We evaluated the tool in a study involving 35 participants. The results were encouraging: the SUS score reached 79.5, and student feedback suggests that the application reduces cognitive effort and improves overall engagement. Many participants found it easier to follow and verify proofs compared to traditional pen-and-paper methods. These findings indicate that \textbf{LogicProof} can serve as a practical and effective supplement to standard teaching approaches.

Future work will focus on extending the range of supported first-order theories. Adding more axiomatic systems would make the tool suitable for more advanced courses in logic and mathematics.

Another important step is integration with Learning Management Systems such as \textit{Moodle}. This would allow automatic grading of assignments and easier result management, reducing the workload for instructors while giving students faster feedback.

The tool is available online \cite{logicProof-app}, and its source code is publicly accessible on GitHub under the MIT license \cite{logicProof-repo}.

\bibliography{bibl}

@inproceedings{karsten2025proofbuddy,
  title={ProofBuddy: How it Started, How it's Going},
  author={Karsten, Nadine and Eiken, Kim Jana and Nestmann, Uwe},
  booktitle = {Proceedings of The 13th International Workshop on Theorem Proving Components for Educational Software (ThEdu24)},
  editor    = {Julien Narboux and Walther Neuper and Pedro Quaresma},
  address   = {Nancy, France},
  month     = {July},
  year      = {2024},
  series    = {Electronic Proceedings in Theoretical Computer Science},
  pages     = {90--111}
}

@article{Korkut2023,
   title    ={A Proof Tree Builder for Sequent Calculus and Hoare Logic},
   volume   ={375},
   ISSN     ={2075-2180},
   DOI      ={10.4204/eptcs.375.5},
   journal  ={Electronic Proceedings in Theoretical Computer Science},
   publisher={Open Publishing Association},
   author   ={Korkut, Joomy},
   year     ={2023},
   pages    ={54–62} 
}

@misc{logika,
   author = {J\'an Perh\'a\v{c}},
   title = {Logic for informaticians: lecture notes},
   year = {2026},
   url = {https://kurzy.kpi.fei.tuke.sk/lpi/},
   note = {(accessed Apr. 30, 2026)}
}

@book{tarski1953,
  title={Undecidable theories},
  author={Tarski, Alfred and Mostowski, Andrzej and Robinson, Raphael Mitchel},
  volume={13},
  year={1953},
  publisher={Elsevier}
}

@misc{coq,
   title = {{The Rocq Proof Assistant, version 9.0.0}},
   autor = {Rocq Development Team, The},
   url = {https://rocq-prover.org/doc/v9.0/refman/index.html},
   note = {Official Release},
   year = {2025}
}

@book{bertot2013,
  title={Interactive theorem proving and program development: Coq’Art: the calculus of inductive constructions},
  author={Bertot, Yves and Cast{\'e}ran, Pierre},
  year={2013},
  publisher={Springer Science \& Business Media}
}

@book{gentzen1969,
  title={The collected papers},
  author={Gentzen, Gerhard},
  volume={166},
  year={1969},
  publisher={North-Holland Publishing Company}
}

@book{zakas2010,
  title={High performance JavaScript: build faster web application interfaces},
  author={Zakas, Nicholas C},
  year={2010},
  publisher={" O'Reilly Media, Inc."}
}

@book{strazzullo2019,
  title={Frameworkless Front-End Development},
  author={Strazzullo, Francesco},
  journal={Berkeley, CA: Apress},
  year={2019},
  publisher={Springer}
}

@book{aho2007compilers,
  title={Compilers Principles, Techniques},
  author={Alfred, V and Monica, S and Ravi, Sethi and Jeffrey D, Ullman and others},
  year={2007},
  publisher={Pearson}
}

@book{huth2004logic,
  title={Logic in Computer Science: Modelling and reasoning about systems},
  author={Huth, Michael and Ryan, Mark},
  year={2004},
  publisher={Cambridge university press}
}

@book{vanDalen2013,
  title={Logic and structure},
  author={Van Dalen, Dirk},
  year={2004},
  publisher={Springer}
}

@misc{sauro2011sus,
  author       = {Jeff Sauro},
  title        = {Measuring Usability with the System Usability Scale (SUS)},
  year         = {2011},
  url          = {https://measuringu.com/sus/},
  note         = {(accessed Apr. 28, 2026)}
}

@misc{rocq-game,
  title        = {Web application - Rocq Game},
  year         = {2026},
  url          = {https://rocq-game.kpi.fei.tuke.sk/},  
  note         = {(accessed Apr. 28, 2026)}
}

@misc{lean-game-server,
  title        = {Lean Game Server},
  note         = {(accessed Apr. 28, 2026)},
  organization = {Mathematisches Institut, Heinrich-Heine-Universität Düsseldorf},
  year         = {2026},
  url          = {https://adam.math.hhu.de/}
}

@inproceedings{perhac2024onlineprover,
  author    = {Ján Perháč and Samuel Novotný and Sergej Chodarev and Joachim Tilsted Kristensen and Lars Tveito and Oleks Shturmov and Michael Kirkedal Thomsen},
  title     = {OnlineProver: Experience with a Visualisation Tool for Teaching Formal Proofs},
  booktitle = {Proceedings of The 13th International Workshop on Theorem Proving Components for Educational Software (ThEdu24)},
  editor    = {Julien Narboux and Walther Neuper and Pedro Quaresma},
  address   = {Nancy, France},
  month     = {July},
  year      = {2024},
  series    = {Electronic Proceedings in Theoretical Computer Science},
  pages     = {56--75}
}

@misc{teachinglogic-dn,
  title        = {Natural Deduction Tool — Teaching Logic Resource},
  note         = {Online resource for natural deduction rules and exercises provided by the Laboratoire d'Informatique de Grenoble, (accessed Apr. 28, 2026)},
  year         = {2026},
  url          = {http://teachinglogic.liglab.fr/DN/}
}

@InProceedings{gasquetEtal2011panda,
  author="Gasquet, Olivier
  and Schwarzentruber, Fran{\c{c}}ois
  and Strecker, Martin",
  editor="Blackburn, Patrick
  and van Ditmarsch, Hans
  and Manzano, Mar{\'i}a
  and Soler-Toscano, Fernando",
  title="Panda: A Proof Assistant in Natural Deduction for All. A Gentzen Style Proof Assistant for Undergraduate Students",
  booktitle="Tools for Teaching Logic",
  year="2011",
  publisher="Springer Berlin Heidelberg",
  pages="85--92",
}

@misc{natural_deduction_tool,
  author = {Stärk, Robert and Lévy, Michel},
  title = {Natural Deduction},
  year = {2023},
  url = {http://teachinglogic.liglab.fr/DN/},
  note = {(accessed Apr. 30, 2026)}
}

@misc{ndprover,
  author  = {{ND-Prover}},
  title   = {Natural Deduction Proof Generator and Checker},
  url     = {https://ndprover.org/},
  note         = {(accessed Apr. 28, 2026)}
}

@misc{logicProof-app,
  title   = {LogicProof: Web application},
  url     = {https://6ximik9.github.io/naturalDeduction/},
  note    = {(accessed Apr. 28, 2026)}
}

@misc{logicProof-repo,
  title        = {LogicProof — Project Repository},
  year         = {2026},
  url          = {https://github.com/6ximik9/naturalDeduction},
  note         = {(accessed Apr. 30, 2026)}
}

@article{miwa2014intelligent,
  title={An intelligent tutoring system with variable levels of instructional support for instructing natural deduction},
  author={Miwa, Kazuhisa and Terai, Hitoshi and Kanzaki, Nana and Nakaike, Ryuichi},
  journal={Information and Media Technologies},
  volume={9},
  number={1},
  pages={132--140},
  year={2014},
  publisher={Information and Media Technologies Editorial Board}
}

@article{sieg2007apros,
  title={The AProS project: Strategic thinking \& computational logic},
  author={Sieg, Wilfried},
  journal={Logic Journal of the IGPL},
  volume={15},
  number={4},
  pages={359--368},
  year={2007},
  publisher={OUP}
}

@phdthesis{perkins2007strategic,
  title={Strategic proof tutoring in logic},
  author={Perkins, Douglas},
  year={2007},
  school={Master's Thesis, Carnegie Mellon}
}

@inproceedings{broda2006friendly,
  title={Friendly e-tutor for Natural Deduction},
  author={Broda, Krysia and Ma, Jiefei and Sinnadurai, Gabrielle and Summers, Alex},
  booktitle={Teaching Formal Methods: Practice and Experience},
  year={2006},
  organization={BCS Learning \& Development}
}

@article{lodder2021generation,
  title={Generation and use of hints and feedback in a Hilbert-style axiomatic proof tutor},
  author={Lodder, Josje and Heeren, Bastiaan and Jeuring, Johan and Neijenhuis, Wendy},
  journal={International Journal of Artificial Intelligence in Education},
  volume={31},
  number={1},
  pages={99--133},
  year={2021},
  publisher={Springer}
}

\end{document}